\documentclass[fleqn,usenatbib]{mnras}

\usepackage[T1]{fontenc}

\DeclareRobustCommand{\VAN}[3]{#2}
\let\VANthebibliography\thebibliography
\def\thebibliography{\DeclareRobustCommand{\VAN}[3]{##3}\VANthebibliography}

\usepackage{graphicx}	
\usepackage{amsmath}	
\usepackage{amssymb}	
\usepackage[table]{xcolor}
\usepackage{multirow}
\usepackage{arydshln}
\usepackage{multirow}
\usepackage{subcaption}

\usepackage{newtxtext,newtxmath} 

\title[Galaxy pairs in S-PLUS DR4]{Structure and large scale environment of galaxy pairs in the S-PLUS DR4}

\author[Cerdosino et al.]{
Cerdosino, M. C.,$^{1}$\thanks{E-mail: candelacerdosino@mi.unc.edu.ar}
O'Mill, A. L.,$^{1,2}$
Rodriguez, F.,$^{1,2}$
Taverna, A.,$^{3}$
Sodré Jr, L.,$^{4}$ 
Telles, E., $^{5}$
\newauthor M\'endez-Hern\'andez, H.,$^{6,7}$
Schoenell, W.,$^{8}$
Ribeiro, T.,$^{9}$
Kanaan, A.,$^{10}$
Mendez de Oliveira, C.$^{4}$\\
\\
$^{1}$ CONICET. Instituto de Astronomía Teórica y Experimental (IATE). Laprida 854, Córdoba X5000BGR, Argentina.\\
$^{2}$ Universidad Nacional de Córdoba (UNC). Observatorio Astronómico de Córdoba (OAC). Laprida 854, Córdoba X5000BGR, Argentina.\\
$^{3}$ Instituto de Astronomía, UNAM, Apdo. Postal 106, Ensenada 22800, B.C., México\\
$^{4}$ Universidade de São Paulo, IAG, Rua do Matão 1226, São Paulo, SP, Brazil\\
$^{5}$ Observatório Nacional, Rua José Cristino, 77 Rio de Janeiro- RJ 20921-400 Brazil\\
$^{6}$ Departamento de Astronom\'ia, Universidad de La Serena, La Serena, Chile. \\
$^{7}$ Instituto de Investigaci\'on Multidisciplinar en Ciencia y Tecnolog\'ia, Universidad de La Serena, La Serena, Chile.\\
$^{8}$ GMTO Corporation 465 N. Halstead Street, Suite 250 Pasadena, CA 91107, USA \\
$^{9}$ Rubin Observatory Project Office, 950 N. Cherry Ave., Tucson, AZ 85719, USA \\
$^{10}$ Departamento de F\'isica, Universidade Federal de Santa Catarina, Florian\'opolis, SC 88040-900, Brazil\\
}

\date{Accepted XXX. Received YYY; in original form ZZZ}

\pubyear{2015}

\begin{document}
\label{firstpage}
\pagerange{\pageref{firstpage}--\pageref{lastpage}}
\maketitle

\begin{abstract}

In this paper, we use photometric data from the S-PLUS DR4 survey to identify isolated galaxy pairs and analyse their characteristics and properties. 
Our results align with previous spectroscopic studies, particularly in luminosity function parameters, suggesting a consistent trait among galaxy systems.
Our findings reveal a high fraction of red galaxies across all samples, irrespective of projected distance, velocity difference, or luminosity ratio. 
We found that the proximity of a neighbour to its central galaxy influences its colour due to environmental effects.
We also found that central and neighbour have different behaviours: central galaxies maintain a stable red colour regardless of luminosity, while neighbour colours vary based on luminosity ratios. When the central is significantly brighter, the neighbour tends to be less red.\
According to our division in red, blue and mixed pairs, we found evidence of galactic conformity.
Red pair fractions increase in closer pairs and in pairs of similar luminosity, indicating shared environments promoting red galaxy formation.\ 
Analysing local density, the expected colour-density relation is of course recovered, but it is strongly determined by the stellar mass of the pair.
In denser environments, the red pair fractions increase, blue pairs decrease and for mixed pairs it depends on their stellar mass: more massive mixed pairs decrease their fraction whereas the lower massive ones increase it.
These results shed light on the intricate relationship between galaxy pairs, their characteristics, and environmental influences on colour, providing insights into their evolutionary histories.

\end{abstract}



\section{Introduction}

It has long been recognized that galaxies are not uniformly distributed throughout the universe \citep[e.g.,][]{Abell1958, Oemler1974}. Instead, they frequently form diverse structures, ranging from pairs of galaxies to large clusters with hundreds of members, covering a wide range of intermediate systems.
Groups of galaxies exhibit a wide range of morphologies, shapes, and sizes, including, for instance, diffuse groups, fossil groups, compact groups, minor systems (such as pairs and triplets), and rich groups \citep[e.g.,][]{Yang2007, OMill2012, Kanagusuku2016, Taverna2016, Rodriguez2020}. 
This means that the definition of a group essentially depends on the number of galaxies and their magnitude distributions, within a particular volume, up to a defined magnitude limit.

In particular, galaxy pairs are the simplest gravitational systems of galaxies, formed by a more massive primary galaxy (the central one) and a less massive secondary galaxy (the neighbour). Since the pioneering works such as \cite{Karachentsev1972}, different authors have studied these systems, generally with different criteria when defining them. 
According to the standard scenario of hierarchical clustering in the universe, larger systems are formed through the accretion of less massive objects \citep[e.g.,][]{Peebles1980,White1991,WhiteRees1978}.
Consequently, we can expect galaxy pairs to behave as the connection between the environments of isolated galaxies and triplets or groups, allowing us to study the galaxy evolution and environmental effects in intermediate-density regions \citep{Ellison2008, Ellison2010, ArgudoFernandez2015, Duplancic2018}.

As is well known, close pairs are valuable laboratories for studying mergers and other processes that drive galaxy evolution \citep[e.g.,][]{Toomre1972,Patton2000,HernandezToledo2006}. These processes can cause significant changes to the structure and evolution of the galaxies involved. By studying these interactions, we can gain a better understanding of how gravity shapes the distribution of stars, gas, and dark matter in galaxies, and how it affects their star formation rates and evolution. 
In this context, pioneering observations indicate that galaxies with close companions exhibit slightly bluer integrated optical colours \citep{Carlberg1994,Patton1997}, suggesting increased star formation activity associated with the proximity of galaxies on scales of a few tens of kiloparsecs.
Studies such as \cite{Lambas2003} and \cite{Ellison2008}, confirmed that the star formation rate is higher in galaxies in pairs than in field galaxies. This can be attributed to early interactions between the members of a pair during the assembly of these systems, which can trigger bursts of star formation \citep{BarnesHernquist1996}.  
While the star formation activity increases as the separation between the members decreases \citep{Barton2000,Lambas2003}, more modest increases are also observed up to 150 kpc \citep{Patton2013}. 
These close pairs are expected to be the progenitors of the merged systems at later times, which by mixing their stellar populations may be the precursors of the red, non-star-forming galaxies observed in dense environments. 
In addition, \cite{Patton2011} found that the red fraction of galaxies in pairs is higher than a control sample, with no clear dependence on the projected separation between the members. 
The authors argued that this effect would not be associated with galaxy interactions or mergers but rather likely with the fact that galaxy pairs reside in higher-density environments compared to unpaired galaxies.
In turn, they detected signs of interaction-induced star formation within only the blue galaxies in the pairs, being strongest in the closest pairs and found mainly in low to medium-density environments. 
When they interpreted it together with a simple model of induced starbursts, their results are consistent with a scenario in which close pericentre steps trigger induced star formation in the centres of galaxies that are sufficiently gas-rich, after which the galaxies gradually redden as they separate and their starbursts age.

Those results are in accordance with the widely known fact that the environment of galaxies affects their properties \citep[e.g.,][]{Dressler1980, Kauffmann2004,Baldry2006,Omill2008}. 
Generally, galaxies that inhabit denser environments are redder, brighter, and have earlier-type morphologies than those in low-density regions.
Different works have studied the environments of galaxy pairs and their interactions, for instance, \cite{Ellison2010}
found that galaxies in the lowest density environments show the largest changes in star formation rate, asymmetry and bluer bulge colours. Their results also show evidence that whilst interactions occur at all densities, triggered star formation is seen only in low-to-intermediate density environments, likely due to the typically higher gas fractions of galaxies in low density environments.  
\cite{ArgudoFernandez2015} discovered that the majority of the pairs belong to the outer regions of filaments, walls, and clusters, displaying general distinctions from galaxies situated in voids. In addition, \cite{Duplancic2020} found that pairs inhabit environments of lesser density compared to triplets and groups and when considering the position within the cosmic web the pairs are associated with void environments. Their results also suggest that differences in the properties of galaxies in pairs (and minor systems) are related to the existence of an extra galaxy member and to the large-scale environment inhabited by the systems.
Therefore these results show that the environment of the pairs also affects their properties.

In exploring the environmental influences on galaxy pairs, particularly regarding their characteristic colours, emerges the phenomenon known as "galactic conformity" \citep{Weinmann2006}. This phenomenon encapsulates the tendency for closeby galaxies to mirror each other's properties, displaying similarities in star formation rates, colours, gas fractions, and morphologies.
This phenomenon was initially observed among satellites orbiting central galaxies \citep{Weinmann2006,Kauffmann2010,Wang2010,Robotham2013}: at fixed group mass, red (blue) central tend to have a redder (bluer) satellite population.  
The study of galactic conformity in galaxy pairs allows us to comprehend how the properties of galaxies are influenced by the close presence of another galaxy, and consequently, how it may influence their evolution.

In order to identify galaxy pairs, the traditional method is based on two quantities that relate the two galaxy members: the projected distance on the sky ($r_p$) and the velocity difference along the line of sight between them ($\Delta V$). By defining a maximum value for these two quantities ($r_{p,max}$ and $\Delta V_{max}$), one can determine whether two galaxies form a pair, as defined here, by checking whether their parameters are below the set maximum. We followed this traditional approach to identify our pairs. 
In the literature, various parameter values for $r_{p,max}$ and $\Delta V_{max}$ can be found since they depend on the catalogue and the specific objective of the study. For instance, works that focus on pair mergers use smaller separations because when $r_p$ and $\Delta V$ between galaxies decrease, the probability of interactions increases \citep[][]{Barton2000,Lambas2003,Alonso2007}. 
On the other hand, papers that aim to study the effects of the presence of a companion opt for less restrictive limits to avoid losing possible real pairs with widely separated members. Hence, separations on the order of megaparsecs and larger velocity differences are commonly employed \citep[][]{Patton2016,NottaleChamaraux2018}.

In general, the velocity difference calculated with spectroscopic redshift information is used for the identification of galaxy pairs. Consequently, most pair catalogues have been mainly constructed using spectroscopic surveys. Nevertheless, nowadays, there are already some works in the literature that were performed with purely photometric surveys \citep{LopezSanJuan2015,Rodriguez2020}.
The last is a consequence of the goal to explore increasingly larger volumes of the universe: state-of-the-art galaxy surveys have embraced a different approach to handle the immense amount of data. Recognising that high-precision spectroscopy may not be the most suitable method for such extensive datasets, these surveys have shifted towards photometric studies combined with photometric redshift estimates of low uncertainty. 
Several projects such as the Physics of the Accelerating Universe Survey \citep[PAUS,][]{Eriksen2019}, the Javalambre-Physics of the Accelerating Universe Survey \citep[J-PAS,][]{Benitez2014}, the Javalambre-Photometric Local Universe Survey \citep[J-PLUS][]{Cenarro2019}, among others, are able to obtain accurate photometric redshifts due to the use of systems with narrow-band filters.
Some of these surveys can achieve redshifts as precise as dz/(1 + z) $\sim$ 0.003 \citep{Alarcon2021,Laur2022,Cenarro2019}.
In particular, the Southern Photometric Local Universe Survey\footnote{\url{https://www.splus.iag.usp.br/}} \citep[S-PLUS,][]{MendesdeOliveira2019} is a purely photometric survey of the southern hemisphere sky, with a filter system consisting of 5 broad-band and 7 narrow-band filters which allows it to obtain a dz/(1 + z) $<$ 0.01 \citep{Lima2022}.

This strategic shift towards photometry enables more efficient data collection while still providing valuable insights into the properties and distribution of galaxies.  
In this scenario, obtaining a reliable sample of galaxy pairs or minor galaxy systems is challenging because of projection and contamination effects, due to the use of photometric data only. Therefore, it is important to apply accurate tests to ensure the recovery of truly bound systems, minimizing losses and contamination.
Works such as \cite{Rodriguez2020} have identified pairs with photometric samples and they found that as photometric redshift errors increase the pairs tend to be misidentified. However they found that overall properties such as the luminosity and mass distributions are successfully reproduced.

The aim of this work is to provide a catalogue of isolated galaxy pairs for the Southern Photometric Local Universe Survey internal Data Release 4. 
We intend to improve the identification codes, taking into account the 12-band photometric system of S-PLUS. We believe that the forthcoming results will significantly enhance the reliability of these group catalogues and the statistical results obtained from them.
In doing so, we aspire to contribute to a more comprehensive characterisation of membership and properties for both the systems and their constituents. This enhanced knowledge will lead to a better understanding of the formation mechanisms and evolution of galaxies in diverse environments. Furthermore, it will contribute to the development of more realistic models of galaxy formation for integration into future semi-analytical models.

The paper is organised as follows. In Section \ref{sec:splus} we describe the survey and data selection. We detail the algorithm for identifying isolated galaxy pairs in Section \ref{sec:pairidentifier} and we test it by constructing a mock catalogue in Section \ref{sec:eval}. The results are described and discussed in Section \ref{sec:results}. Finally, a brief summary and conclusions are given in Section \ref{sec:conclusions}.
For this work, we assume the standard $\Lambda$CDM cosmology of \citet{Planck16} with $\Omega_{\rm m}$ = 0.3089, $\Omega_{\Lambda}$ = 0.6911 and $h$= 0.6774.

\section{DATA: S-PLUS Galaxy survey} \label{sec:splus}

The Southern Photometric Local Universe Survey (S-PLUS) is an imaging survey that plans to cover $\sim$9300 deg$^2$ of the southern hemisphere sky in the optical range, using a robotic telescope located at Cerro Tololo Interamerican Observatory (CTIO), Chile.
In this work, we use the observations of the internal S-PLUS Data Release 4 in the STRIPE 82 region (rectangular area within the coordinates $0^{\circ}<RA<60^{\circ}$, $300^{\circ}<RA<360^{\circ}$ and $-1.4^{\circ}<DEC<+1.4^{\circ}$), and we adopted the Petrosian magnitudes corrected by extinction \citep{1998ApJ...500..525S}.
These observations were made with the 12 S-PLUS/Javalambre optical filter system: 5 broadband filters similar to the Sloan Digital Sky Survey \citep[SDSS,][]{York2000}: $u$, $g$, $r$, $i$, $z$, and 7 narrow-band filters. This filter system is ideal for a better photometric redshift estimation of galaxies in the nearby universe \citep{Cenarro2019}.  
S-PLUS provides photometric redshifts both as point estimates and as probability distribution functions for all objects. The photometric redshifts are obtained using a Bayesian Mixture Density Network model \citep{10.1063/1.1144830,10.1162/089976699300016728}, which is a supervised machine learning algorithm. More details about the method can be found in \citet{Lima2022}.

\subsection{Improving the accuracy of galaxy sample}

Photometric classification is provided for all sources in DR4, including QSO/star/galaxy probabilities \citep{2021MNRAS.507.5847N}. The classification is performed by two Random Forest algorithms that were re-trained with DR4 photometric information of objects classified from spectroscopy from SDSS. However, in this work, in order to improve this classification, we have applied additional constraints to the data by using surveys that have independent classifications and share area with S-PLUS DR4, like Gaia and SDSS. 
To do so, we first cross-match all the S-PLUS DR4 objects with sources in Gaia DR3 catalogue \citep{Gaia2022b}, within a search radius of $10^{-6}$ degrees. 
We then cleaned the stars from the sample up to the Petrosian magnitude $m_r=18$, where Gaia DR3 is complete.
For this, we used the parameter RUWE \citep[Renormalised Unit Weight Error;][]{Lindegren2018} which is a statistical indicator that can be used to assess the quality and reliability of astrometric data.
The RUWE is expected to be around 1.0 for sources where the single-star model provides a good fit to the astrometric observations. Additionally, a value greater than 1.4 could indicate that the source is non-single or otherwise problematic for the astrometric solution. Therefore, we considered stars those objects with $1 <$ RUWE $< 1.4$ \citep{Berger2020} and removed them from the S-PLUS catalogue.
After this, we made a second cross-match with the SDSS DR17 catalogue \citep{Abdurrouf2022} and used its star/galaxy classification to remove from our galaxy catalogue those objects classified as stars for magnitudes higher than 18.
Although the surveys we utilised cover similar areas, S-PLUS is deeper and contains objects that are absent in the other surveys. When an object is not found in either Gaia or SDSS, we assign to it the S-PLUS classification.

\subsubsection{Galaxy sample}

Our final sample consists of galaxies on Stripe 82 with Petrosian magnitudes corrected for extinction and photometric redshifts. Photometric redshift measurements can be subject to significant uncertainties, arising from both random and systematic errors. However, as demonstrated by \citet{Lima2022}, narrow-band filters employed in surveys like S-PLUS enable the acquisition of more accurate photometric redshifts. Additionally, deep learning models have exhibited superior performance in estimating photometric redshifts compared to traditional methods. In this study, we further minimize uncertainties by focusing on pairs of bright galaxies. To establish the most complete and reliable sample of galaxy pairs, we opted to utilize two galaxy samples, defined by the signal-to-noise ratio (S/N) of the galaxy in the detection image. This relationship, as defined in \citet{2022MNRAS.511.4590A}, represents the flux of the galaxy relative to its flux measurement error.

The first sample, known as the SN10 sample, comprises galaxies with a signal-to-noise ratio (S/N) greater than 10. This sample has higher purity and low contamination, but it is smaller than the second sample, known as the SN5 sample, comprising galaxies with a S/N  greater than 5. This sample is larger than the SN10 sample, but it has lower purity and may be contaminated by spurious pairs.

\begin{figure}
\centering
	\includegraphics[width=0.9\columnwidth]{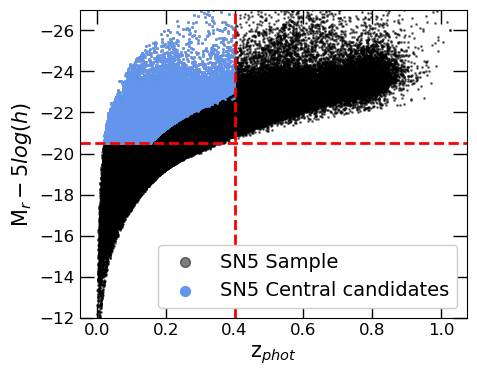}
    \caption{Absolute magnitude ($M_r-5log(h)$) vs. photometric redshift ($z_{phot}$) for the S-PLUS DR4 SN5 sample. In black points, it shows the complete flux sample (14 $\leq m_r \leq$ 20.8) of S-PLUS SN5 galaxies, and in blue the SN5 central galaxy candidates. The red dashed lines show the redshift and absolute magnitude limits of a volume complete sample of central galaxies. }
    \label{fig:mvsz}
\end{figure}

For the purpose of obtaining complete flux samples, we used Petrosian $r$-band apparent magnitudes smaller than $m_{r,lim}$ = 20.2 for SN10 and $m_{r,lim}$ = 20.8 for SN5 and took as a minimum value $m_r$ = 14 (\citealt{Costa-Duarte2019}).
The number of objects in SN10 and SN5 samples are 410828 and 822641, respectively.
Figure \ref{fig:mvsz} shows in black the $r$-band absolute magnitude ($M_r-5log(h)$) as a function of photometric redshift ($z_{phot}$) for the complete sample in galaxy flux. 
The central galaxy candidates are shown in blue, and we will explain them in Section \ref{sec:pairidentifier}. 
To compare the photometric properties of galaxies at different redshifts, their magnitudes must be corrected for the changes in effective rest-frame wavelengths of filter bandpasses, known as K-corrections. For this reason, we employed the publicly available software package of \citet{Blanton2007} in version $V4\_3$ to obtain de-reddened model magnitudes at $z = 0.$

\section{Galaxy pairs identification} \label{sec:pairidentifier}

In order to obtain a catalogue of galaxy pairs, we implemented an algorithm to identify them. In this section, we describe the methodology used to identify the pairs of galaxies and in the next Section how we tested it.

\subsection{Implementation of the identification algorithm} \label{sec:identifier}

As mentioned above, we decided to follow a traditional approach to galaxy pair identification using $r_{p,max}$ and $\Delta V_{max}$ between galaxy members, but taking into account the uncertainties in the photometric redshifts. 
The algorithm of identification was developed based on \cite{OMill2012} to identify galaxy triplets in SDSS DR7 \citep{Abazajian2009} with photometric and spectroscopic redshift and on \cite{Rodriguez2020} to identify close pairs in the PAUS catalogue with photometric redshift. In this work, we introduced some modifications to these codes to adapt them to the S-PLUS observational catalogue, as well as other improvements in the identification.

The identification algorithm can be summarised in 3 steps: 

\begin{enumerate}
    
\item As a first step, the algorithm searches for candidate central galaxies, where the ``central galaxy'' is defined as the most luminous one of the pair. We considered as candidates those with a Petrosian $r$-band absolute magnitude $M_{r,c} -5log(h) \leq$ -20.5, $z \leq$ $z_{lim}$ and a $r$-band apparent magnitude two magnitudes lower than the limiting catalogue magnitude, i.e. $ m_{r,c} \leq m_{r,lim} - 2 $. The chosen values define a volume-complete sample for the central candidates and they are $m_{r,lim} = 20.2$ and $z_{lim}=0.3$ for SN10, and $m_{r,lim} =20.8$ and $z_{lim}=0.4$ for SN5. Table \ref{tab:samples} summarises these values for both samples. 
To illustrate this, Figure \ref{fig:mvsz} shows the case of sample SN5. In black is shown the absolute magnitude of the galaxies in the sample as a function of the photometric redshift, with the central galaxy candidates highlighted in blue.

The condition of $ m_{r,c} \leq m_{r,lim} - 2 $ is added to guarantee that the other galaxy in the pair is not outside the sample limiting magnitude and that we do not lose pairs due to it; for example, \cite{DiazGimenez2010} used a similar criterion for the detection of compact groups, with 3 magnitudes instead of 2.

\begin{table}
\begin{center}
\begin{tabular}{|c|c|c|c|c|}
\hline
Sample & $m_{r,lim}$ & $m_{r,c}$ & $M_{r,c} -5log(h)$ & $z_{lim}$ \\ \hline
SN10 & 20.2 & 18.2 & -20.5 & 0.3 \\
SN5 & 20.8 & 18.8 & -20.5 & 0.4 \\ \hline
\end{tabular}
\caption{Sample values. $m_{r,lim}$ is the limiting Petrosian $r$-band apparent magnitude of the sample; $m_{r,c}$, $M_{r,c}$ and $z_{lim}$ are the limiting $r$-band apparent magnitude, $r$-band absolute magnitude and redshift of the central candidates, respectively. }
\label{tab:samples}
\end{center}
\end{table}

\item Secondly, the algorithm searches for each central galaxy candidate for its fainter neighbours within a projected distance and a velocity difference smaller than the defined limits (i.e. $r_p \leq r_{p,max}$ and $\Delta V \leq \Delta V_{max}$). 
We added a $r_{p,min}$ = 10 $h^{-1}$ kpc, such that $r_{p,min} \leq r_p$, to avoid misidentifications \citep{Ellison2008}. 
In addition, we established a limiting maximum apparent magnitude difference of 2 magnitudes ($\Delta m_r = 2$) between them. This last condition, together with the luminosity limit of the central galaxies, ensures the identification of real pairs and not the case of a luminous galaxy with a dwarf satellite or dwarf galaxy (e.g. \citealt{SalesLambas2005}). 
Limiting the apparent magnitude difference also ensures galaxies of similar mass. Then, we consider as pairs those central galaxies that have only 1 neighbour and no other central galaxy that satisfies the conditions mentioned above.

\item Finally, we applied an isolation criterion to ensure that the pairs are not part of a larger system. This is done by checking that there are no other galaxies within a projected distance between $r_{p,max}$ and $3 \times r_{p,max}$, and a velocity difference of $\Delta V < \Delta V_{max}$. It is important to note that this isolation criterion does not imply that the system is totally solitary. It could be, for example, on the edge of a more massive system.\\
\end{enumerate}

The choice of the values of $r_{p,max}$ and $\Delta V_{max}$ will be those that allow better and safer identification of the pairs. To evaluate this, we performed purity and completeness tests using a simulated catalogue that follows the physical properties of S-PLUS, which is analysed in the next section.

\section{Evaluation of the galaxy pair identifier} \label{sec:eval}

\subsection{The mock catalogue} \label{sec:mock}

In order to test the proposed identification procedure we built a mock catalogue with the photometric characteristics of the S-PLUS DR4. The aim is to predict the observational results and, with this information, a) evaluate the assignment made by the identifier and b) estimate the errors that should be taken into account when applying the method to the real catalogue.
For this purpose, we used the galaxy and dark-matter halo catalogues from {\it The Next Generation} Illustris (IllustrisTNG, \citealt{Nelson2019}) magneto-hydrodynamical cosmological simulations, which represent an updated version of the Illustris simulations \citep[][]{Vogelsberger2014a,Vogelsberger2014b,Genel2014}. They are performed with the arepo moving mesh code \citep{Springel2010} and include sub-grid models that account for radiative metal-line gas cooling, star formation, chemical enrichment from SNII, SNIa, and AGB stars, stellar feedback, supermassive black hole formation with multimode quasar, and kinetic black hole feedback. The main updates to this simulation are a new implementation of black hole kinetic feedback at low accretion rates, a revised scheme for galactic winds, and the inclusion of magneto-hydrodynamics \citep[][]{Pillepich2018, Weinberger2017}.

We used the IllustrisTNG300-1 run (hereafter TNG300), the largest simulated box from the IllustrisTNG. This run adopts a cubic box of side $205\,h^{-1}$~Mpc with periodic boundary conditions. The TNG300 run follows the evolution of 2500$^3$ dark-matter particles of mass $4.0 \times 10^7 h^{-1} {\rm M_{\odot}}$, and 2500$^3$ gas cells of mass $7.6 \times 10^6 h^{-1} {\rm M_{\odot}}$. 
Dark matter halos in TNG300 are identified using a friends-of-friends (FOF) algorithm with a linking length of 0.2 times the mean inter-particle separation \citep{Davis1985}. 
Subhaloes are afterwards identified using the SUBFIND algorithm \citep{Springel2001, Dolag2009}, and those containing a stellar component are considered galaxies. Typically, each dark-matter halo contains multiple galaxies, including a central galaxy and several satellites.

To build our mock catalogue, we followed a procedure used in \cite{Rodriguez2015}, implemented in  \cite{RodriguezMerchan2020} and  \cite{Rodriguez2021}. First of all, we placed the observer at the origin of the TNG300 box. Considering the box periodicity, we simulated the volume of the S-PLUS catalogue by adding the TNG300 volume repeatedly. The spatial resolution of this simulation is suitable for the implementation of this algorithm. The galaxy redshifts were calculated by combining the cosmological distance and the distortion produced by proper motions. Magnitudes based on the summed-up luminosities of all the stellar particles provided by the simulation are used, in particular, $r$-band. 
From these redshifts and magnitudes, we derived the apparent magnitudes of each galaxy.  
To construct the catalogue, we used a volume that simulates the same angular area (in the STRIPE 82 region) and depth as the S-PLUS. To mimic the S-PLUS limited flux selection, we applied the same apparent magnitude threshold ($m_r$= 20.8 and 20.2 in $r$-band for the SN5 and SN10 samples, respectively) as the S-PLUS.
For this, photometric redshifts ($z_{ph}$) are calculated by adding an uncertainty that imitates the errors of the photometric measurements ($z_{err}$) to the galaxy redshift ($z$): $z_{ph} = z + z_{err}$. The $z_{err}$ values are calculated for each galaxy by choosing a random value whose probability is given by a Gaussian distribution with a standard deviation determined by the S-PLUS error $\sim 0.01 (1+z)$, as estimated for low-redshift galaxies by \cite{Lima2022}. Thus, the error distribution of the simulated catalogue reproduces the estimates in the observations. Through this procedure, we obtained a  galaxy catalogue with right ascension, declination, apparent magnitudes, halo mass, redshift and photometric redshift. We verified that the apparent magnitude distributions match with the observed data. In addition, the number counts for different magnitude cuts are reproduced.

Finally, to evaluate our pair identification method, we implemented the procedure described in Section \ref{sec:identifier} on the mock catalogue. We assessed its performance using several parameters in order to get an optimal identification of galaxy pairs in S-PLUS.

\subsection{The algorithm evaluation: Completeness and Contamination test} \label{sec:pyc}

The major source of contamination in the detection of photometric systems is due to the error in photometric redshifts. This is because the uncertainties in photometric redshifts affect the choice of $\Delta V$ within which to search for galaxy pairs.
In order to evaluate our finder performance, we used the mock galaxy redshift surveys described in the previous section to estimate the expected completeness and contamination of our pair samples.

The completeness is defined as $N_{rec}/N_r$, where $N_{rec}$ is the number of pairs candidates identified with our algorithm that is matched to the true pairs in the mock (recovered pairs) and $Nr$ is the number of true pairs in the mock (real pairs) within the same halo. 
On the other hand, the contamination is defined as $N_{sp}/N_r$, where $N_{sp}$ is the number of systems detected whose member galaxies do not belong to the same halo (spurious pairs).

\subsubsection{Selection of the identifier parameters} 
\label{sec:pyc_values}

\begin{figure*}
\centering
	\includegraphics[width=1.4\columnwidth]{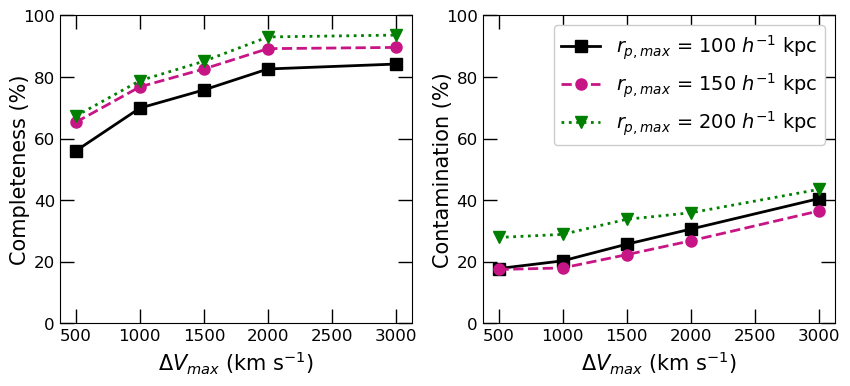}
    \caption{Mean values, expressed in percentage, of the completeness (\textit{left panel}) and contamination (\textit{right panel}) of the pairs obtained for different $r_{p,max}$ and $\Delta V_{max}$, for the SN10 sample.} 
    \label{fig:pyc}
\end{figure*}

To find the best parameters for pair identification, we tested different values of $r_{p,max}$ and $\Delta V_{max}$. We considered $r_{p,max}$ = 100, 150, and 200 $h^{-1}$ kpc, and for each of these values, we varied $\Delta V_{max}$ between 500 and 3000 $~\mathrm{km\,s}^{-1}$. 
Figure \ref{fig:pyc} shows the completeness and contamination as a function of $\Delta V_{max}$, with different $r_{p,max}$ values shown in different colours. We show only the results for the SN10 sample, as the results for the SN5 sample were not significantly different. This plot helps us to find the best compromise between completeness and contamination.
Taking into account this compromise, we decided to use $r_{p,max}$ =150 $h^{-1}$ kpc and $\Delta V_{max}$ =1000 $~\mathrm{km\,s}^{-1}$ for the identification of our pairs. With these values, we obtained a completeness of $\sim 77\%$ and a contamination of less than $\sim 20\%$. 

The limits on $r_{p,max}$ and $\Delta V_{max}$ chosen are not too restrictive to avoid losing possible real pairs with widely separated members and taking into account the uncertainties introduced by photometric redshifts. This selection also avoids reducing the sample size too much, which can be a limiting factor in pair statistics. However, the values are also not too high, so as to not depart from those found in the literature and to select systems that are gravitationally bound.

\section{Analysis and Results} \label{sec:results}

\begin{figure*}
    \centering
    \begin{subfigure}[!b]{0.28\textwidth}
        \includegraphics[width=\textwidth]
        {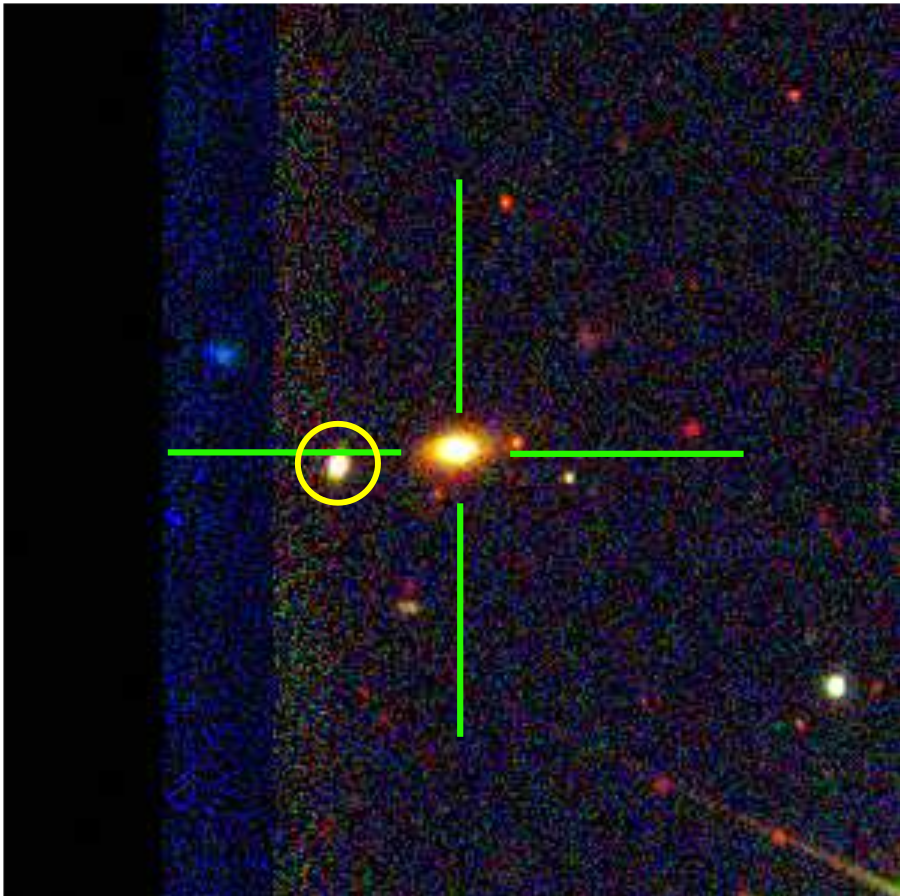} 
    \end{subfigure}
    \qquad
    \begin{subfigure}[!b]{0.28\textwidth}
        \includegraphics[width=\textwidth]
        {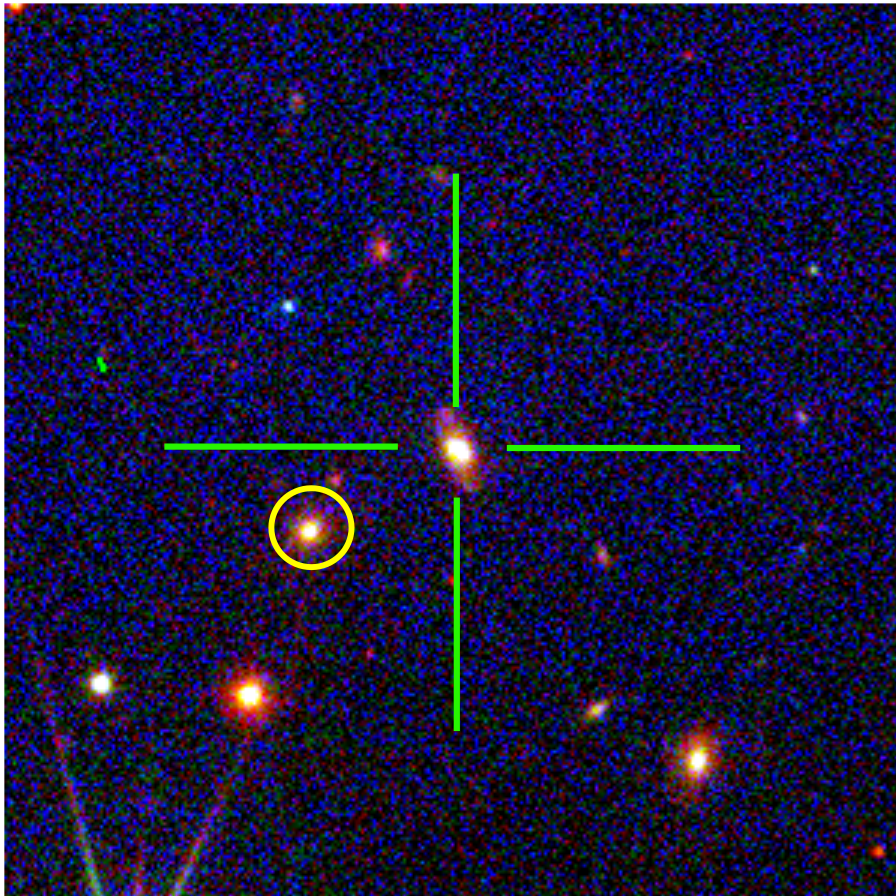}
    \end{subfigure}
    \qquad
    \begin{subfigure}[!b]{0.28\textwidth}
        \includegraphics[width=\textwidth]
        {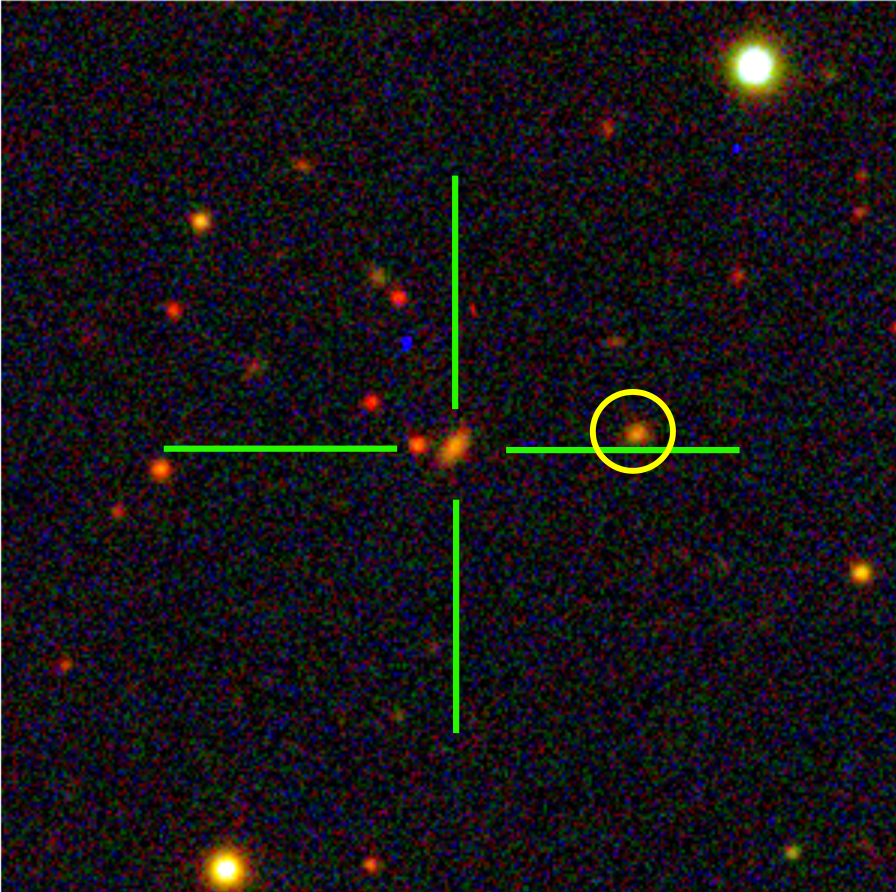}
    \end{subfigure}
\caption{Examples of S-PLUS galaxy pairs, obtained from 12 band combination, 300 pixels per side. The pairs are centered on the central galaxy and the yellow circle corresponds to the companion galaxy. \textit{Left panel}: $z_{phot} = 0.051$. \textit{Middle panel}: $z_{phot}=0.13$. \textit{Right panel}: $z_{phot}=0.25$. }
\label{fig:ejpairplus}
\end{figure*}

We applied the algorithm with the parameters obtained from the completeness and contamination analysis ($r_{p,max}$ = 150 $~\mathrm{kpc}$ and $\Delta V_{max}$ = 1000 $~\mathrm{km\,s}^{-1}$, see \ref{sec:pyc_values}) to our samples. We obtained 1278 and 700 isolated galaxy pairs for SN5 and SN10, respectively.
In Figure \ref{fig:ejpairplus}, we show 3 examples of our SN10 galaxy pairs at different redshifts and centered on the central galaxy. 
The images were downloaded from the S-PLUS website which offers tools for visualising the fields\footnote{\url{https://splus.cloud/imagetools/12filter}}. They were created with the Trilogy\footnote{\url{https://www.stsci.edu/~dcoe/trilogy/Intro.html}} tool by combining data from the 12 S-PLUS bands and 300 pixels on a side. 

With the samples obtained, we studied in the next subsections the colours, stellar masses, and environments of the pairs and how they relate to each other.

\subsection{Properties of galaxy pairs}

To analyse the behaviour of both SN10 and SN5 galaxy pair samples, we have chosen two colours, gold and black, respectively, to identify them in the following figures. Figure \ref{fig:pares} shows the distributions of the photometric redshift of the central and neighbour galaxies, the velocity difference between members of a pair, and the luminosity ratio between the neighbour and central galaxies. We also added in the figure the fraction of pairs as a function of the projected distance between the members of the pairs.

The top left panel shows the photometric redshift \textbf{($z_{phot}$)} distributions of the central galaxies (solid line, $z_c$) and the neighbour galaxies (dashed line, $z_n$). 
It can be seen that both central and neighbour galaxies have the same distribution for each of the samples. This was expected since we are looking for gravitationally bound systems and therefore their galaxies should have similar redshifts. 
It is highlighted that the mean value of the $z_{phot}$ distributions are 0.14 and 0.17 for SN10 and SN5, respectively. 
As can be seen, the SN5 sample is slightly shifted towards higher redshifts because it has a higher redshift limit.

In the top right panel of Fig. \ref{fig:pares}, we show the pair fraction as a function of their projected distance ($r_p$) between pair members. We can see that the fraction increases with $r_p$, indicating that there are more pairs with larger projected distances.  
In the bottom left panel, we show the velocity difference distribution ($\Delta V$) between the pair members. The two samples exhibit consistent behaviour, with a slight dip at $\sim 500 ~\mathrm{km\,s}^{-1}$, probably due to statistical fluctuations (at two sigma and three sigma levels for SN10 and SN5, respectively).

Finally, in the bottom right panel, we can see the distribution of the ratio between luminosity in the $r$-band of the neighbour galaxy and the central galaxy ($Lr_n/Lr_c$).
The distributions have a peak at $Lr_n/Lr_c \sim 0.2$ and are consistent with each other with a median of $\sim 0.43$.
When we focus on the first and last segments of the distribution, we notice that $36\%$ ($38\%$) of SN10 (SN5) pairs exhibit their brightest central galaxy as the dominant member with $Lr_n/Lr_c < 0.3$. In contrast, $15\%$ of both SN10 and SN5 pairs have member galaxies with comparable luminosities ($Lr_n/Lr_c > 0.7$).
This decreasing behaviour of the luminosity ratio distribution is characteristic of the pairs and has been observed in other samples \citep[e.g.,][]{Lambas2012,Gonzalez2023}.

\begin{figure*}
\centering
	\includegraphics[width=1.7\columnwidth]{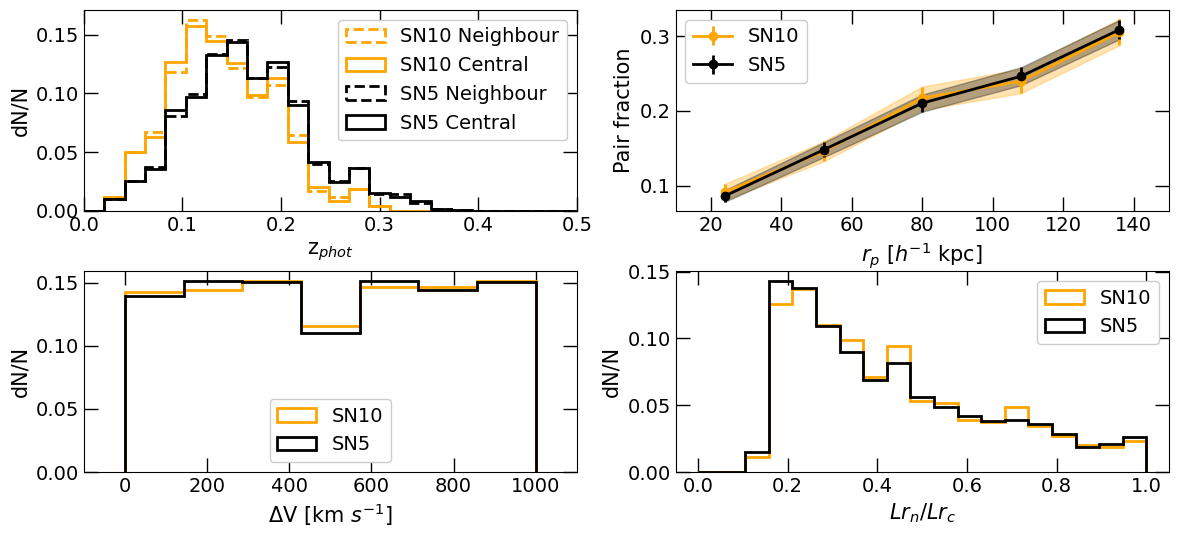} 
    \caption{ Isolated galaxy pairs: SN10 sample in gold and SN5 sample in black.  
    \textit{Left top panel}: photometric redshift ($z_{phot}$) normalized distribution of central (solid line) and neighbour (dashed line) galaxies. \textit{Right top panel}: pair fraction as a function of projected distance ($r_p$). \textit{Left bottom panel}: velocity difference ($\Delta V = |z_c-z_n|c$) normalized distribution between pair members. \textit{Right bottom panel}: normalized distribution of the $r$-band luminosity ratio between the neighbour and central galaxy ($Lr_n/Lr_c$). 
    } 
    \label{fig:pares}
\end{figure*}

\subsection{Stellar mass and Luminosity Function} \label{mass}

One of the strengths of our S-PLUS DR4 sample is that accurately measured colours are available for the galaxies. 
This gives us the possibility of  investigating different properties of our galaxy pairs, such as their stellar mass, determined through photometric techniques, and the luminosity function.

Analysing the stellar mass distribution within small galaxy groups is crucial for comprehending the processes of galaxy formation and evolution. 
The definition of a "small galaxy group" is not universally agreed upon, leading to variations in the literature. For instance, studies by \citet{2007ApJ...671..153Y} and \citet{article} examine different properties, including stellar mass, for groups ranging from a single isolated galaxy, galaxy pairs, triplets, and groups with more than four members.
Following \citet{2009ApJ...695..900Y} and \citet{2007ApJ...671..153Y}, we used the relation between stellar mass-to-light ratio and colour of \cite{Bell2003} to estimate the stellar mass (M$_*$) of our galaxies. It is defined as a function of the ($M_g-M_r$) colour and the $r$-band absolute magnitude, as follows:
 
\begin{equation}
\begin{split}
\log \left[\frac{M_*}{h^{-2}M_{\odot}}\right] & = -0.306 + 1.097(M_g - M_r)- 0.1 \\
& - 0.4(M_r - 5 \log h - 4.64) 
\end{split}
\end{equation}

\begin{figure*}
    \centering
    \includegraphics[width=1.9\columnwidth]{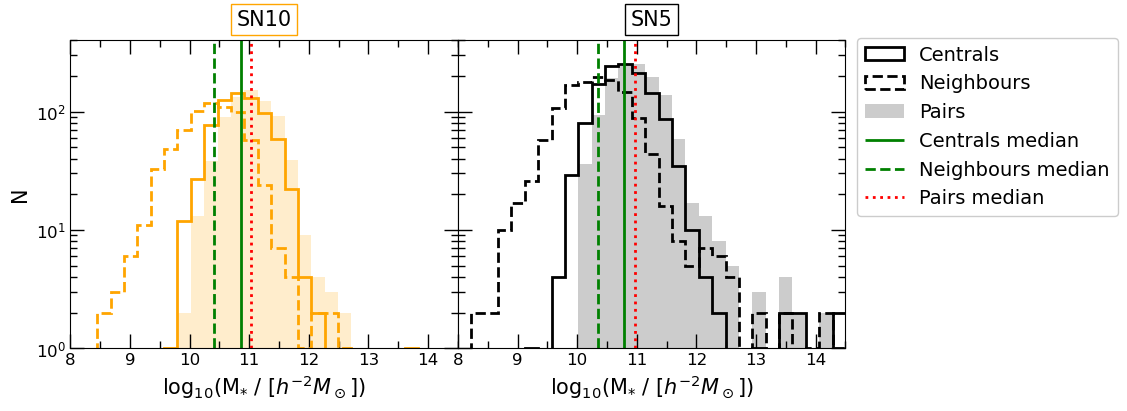}
    \caption{Stellar mass (M$_*$) distributions of SN10 (gold, \textit{left panel}) and SN5 (black, \textit{right panel}) pairs. Solid lines show the distribution for central and dashed lines for neighbour galaxies. Shaded histograms show the total stellar mass of the pairs. Vertical lines indicate the median of the central (green solid), neighbour (red dashed), and total (red dotted) populations. }
    \label{fig:masas}
\end{figure*}

In the left panel of Figure \ref{fig:masas}, we show the calculated stellar masses for the SN10 pairs and in the right panel for the SN5 pairs.
We distinguished between the masses of the central (solid lines) and neighbour (dashed lines) galaxies, where can be seen, as expected, that the central ones are more massive. The medians of the central and neighbours stellar masses are indicated in green vertical lines, which shows a difference in log$_{10}$ of 0.45 dex in both samples. 
We also added the total stellar mass of the pairs in the shaded histograms, which are in the range $10 \lesssim \log_{10}(M_*/h^{-2}M_{\odot}) \lesssim 13$.
The red vertical line shows the median value for the galaxy pairs (11.02 and 10.97 for SN10 and SN5, respectively).  

The range of stellar masses obtained for our galaxies in pairs is similar to that of other authors who analyze pairs of galaxies. For instance, \citet{Robotham2014} conducted an analysis of close galaxy pairs using data from the Galaxy And Mass Assembly II (GAMA-II) redshift sample. These authors obtain a stellar mass range for the galaxy in pairs of the order of 8-13 $\log_{10}(M_*/h^{-2}M_{\odot})$. They investigated mergers of these galaxy pairs and proposed that these mergers have a quantifiable impact on the stellar mass function of galaxies, predicting an increase in characteristic stellar mass of up to $0.01$-$0.05$ dex.
\citet{Patton2016} studied pairs of galaxies in the Sloan Digital Sky Survey (SDSS) with varying projected separations and the influence of the partner on their properties, such as stellar mass, finding ranges similar to those obtained in this work for our projected distances: 8-12 $\log_{10}(M_*/h^{-2}M_{\odot})$. \\

On the other hand, we calculated the luminosity function (LF) of the galaxies in pairs in the $r$-band. We employed the 1/$V_{max}$ method \citep{Schmidt1968} to determine the luminosity function, taking into account incompleteness through a $V$/$V_{max}$ test. This method takes into account the volume of the survey enclosed by the galaxy's redshift, as well as the difference between the maximum and minimum volumes in which it can be observed.
We used the Schechter Function \citep{schecter} to fit the luminosity function. This function describes the spatial density of galaxies as a function of their luminosity. The Schechter function has three parameters: characteristic absolute magnitude ($M^*$), which denotes the 'knee' of the function; the slope ($\alpha$), which represents the power law dominating the faint-end; and the characteristic density ($\Phi^*$):

\begin{equation}
\begin{split}
\Phi(M_r) dM_r = 0.4 ln(10)\Phi^* (10^{0.4(M^*-M_r)})^{(\alpha+1)} \\ 
 \exp(-10^{0.4(M^* - M_r)}) dM_r
\end{split}
\end{equation}

\begin{figure}
    \centering
    \includegraphics[width=0.9\columnwidth]{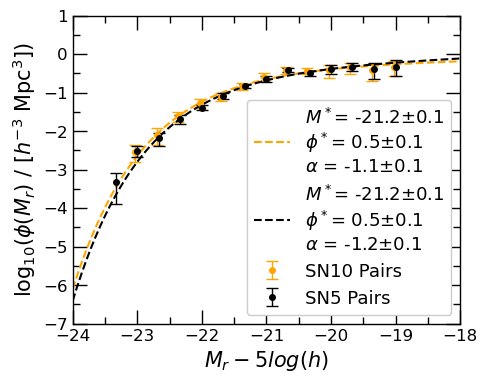}
    \caption{Luminosity function of galaxies in pairs for SN10 (gold points) and SN5 (black points). The dashed lines show the best fit of the Schechter function.}
    \label{fig:lf}
\end{figure}

The luminosity function is shown in Figure \ref{fig:lf}, where dashed lines show the Schechter function fit. The errors in the luminosity function were computed using Jackknife technique. We can see that the parameters obtained are virtually identical for both samples.

Our results for the characteristic luminosity ($M^* =-21.2 \pm 0.1$) and faint-end slope ($\alpha=-1.1 \pm 0.1$) of galaxy pairs are consistent with those obtained by other authors who used the SDSS photometric filter in the $r$-band. For example, \citet{2019MNRAS.485.4474O} obtained similar parameters for low X-ray luminosity poor galaxy clusters.
In addition, \citet{2011MNRAS.415.2553Z} obtained the Schechter parameters for luminosity functions within galaxy groups at low densities regions and redshift up to 0.2. They found that these parameters vary as a function of group mass, with characteristic magnitudes ranging from -20.6 to -20.8 mag and faint-end slopes from -0.85 to -1.1. 
Although the properties of galaxy pairs, small groups, and loose or poor galaxy clusters may differ, the parameters of their LFs are remarkably similar. This may suggest that the luminosity function with parameters similar to those reported here is a characteristic of galaxy systems.

\subsection{Characterisation of the galaxy colours in pairs}

In order to carry out a photometric analysis of the colours of the pairs, we established a criterion to classify the galaxies within the pairs as either blue or red.

\subsubsection{Red and blue galaxies definition}  \label{sec:colour}

\begin{figure}
\centering
	\includegraphics[width=0.9\columnwidth]{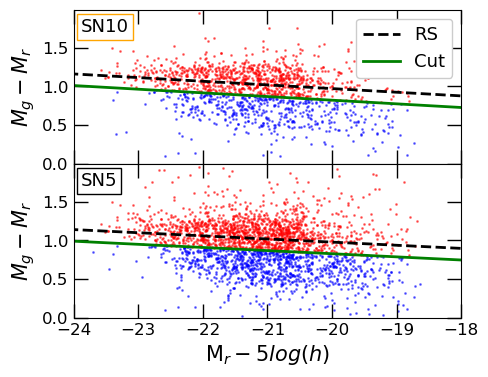}
    \caption{Colour-magnitude diagram for the galaxies in isolated pairs for SN10 (\textit{top panel}) and SN5 (\textit{bottom panel}) samples. The black dashed line shows the red sequence (RS) best fit and the green solid line shows our colour cut between red and blue galaxies.} 
    \label{fig:ajuste}
\end{figure}

Following the methodologies outlined by \cite{Cassata2008} for galaxies and \cite{Balogh2009} for galaxy groups, we proceeded to calculate the red sequence for our sample of galaxy pairs. 
To do this, we analysed the colour distribution ($M_g-M_r$) as a function of $r$-band absolute magnitude of the full sample of pairs, as shown in Figure \ref{fig:ajuste}. 
As a consequence of the magnitude cut applied in constructing the pair sample, the central galaxies and their neighbours reside in distinct volumes and, for that reason, we used a 1/$V_{max}$ weight to construct a statistically complete volume-limited sample within our redshift range.
However, the $V_{max}$ corrections we obtained were very small, because we are considering gravitationally bound galaxies which have similar volumes.

To establish the red sequence, we selected galaxies with $0.7 < M_g-M_r < 1.4$ and $-23.5 < M_r < -19$, and calculated the $V_{max}$-weighted mean colour in bins of $r$-band absolute magnitude. The weighted and unweighted mean values are indistinguishable within their uncertainties. 
For the $V_{max}$-weighted mean colour values, we fitted a straight line and we considered the resulting line as the red sequence.
We show the red sequence (RS) as the black dashed line in Fig. \ref{fig:ajuste} for both samples. We obtained the fit $a=-0.047 \pm 0.003$ and $b=0.02 \pm 0.07$ for SN10, and $a=-0.041 \pm 0.005$ and $b=0.2 \pm 0.1$ for SN5, where $a$ is the slope and $b$ is the y-intercept of the line.

Following \citet{Balogh2009}, we assumed a separation between red and blue galaxies at 0.15 magnitudes below the red sequence.
The solid green line in Fig. \ref{fig:ajuste} represents the colour cut between the two populations. Once the separation was defined, we applied it to the pair samples and obtained 908 (64.8 $\%$) red galaxies and 492 (35.1 $\%$) blue galaxies for SN10, and 1539 (60.2 $\%$) red and 1017 (39.8 $\%$) blue for SN5. Fig. \ref{fig:ajuste} shows in red and blue colours both populations and Table \ref{tab:pares} summarises the numbers of galaxies obtained in each sample.

\subsubsection{Red galaxy fraction}

Once we categorised galaxies into blue and red, we investigated the relationships between different properties of galaxies in pairs with the colour $M_g-M_r$. Our chosen colour division criterion reveals a significantly higher proportion of red galaxies in our samples, as evident in Table \ref{tab:pares}. This aligns with other observational and with simulations studies \citep[e.g.,][]{Barton2007,Perez2009}, who also reported redder galaxies in pairs than their isolated counterparts. In particular, \cite{Patton2011} found an increased red fraction in paired galaxies compared to a control sample matched in stellar mass and redshift, at all projected distances that they considered. 
With this in mind, we examined how these fractions are related between our central and neighbour galaxies.

\begin{figure*}
\centering
	\includegraphics[width=1.5\columnwidth]{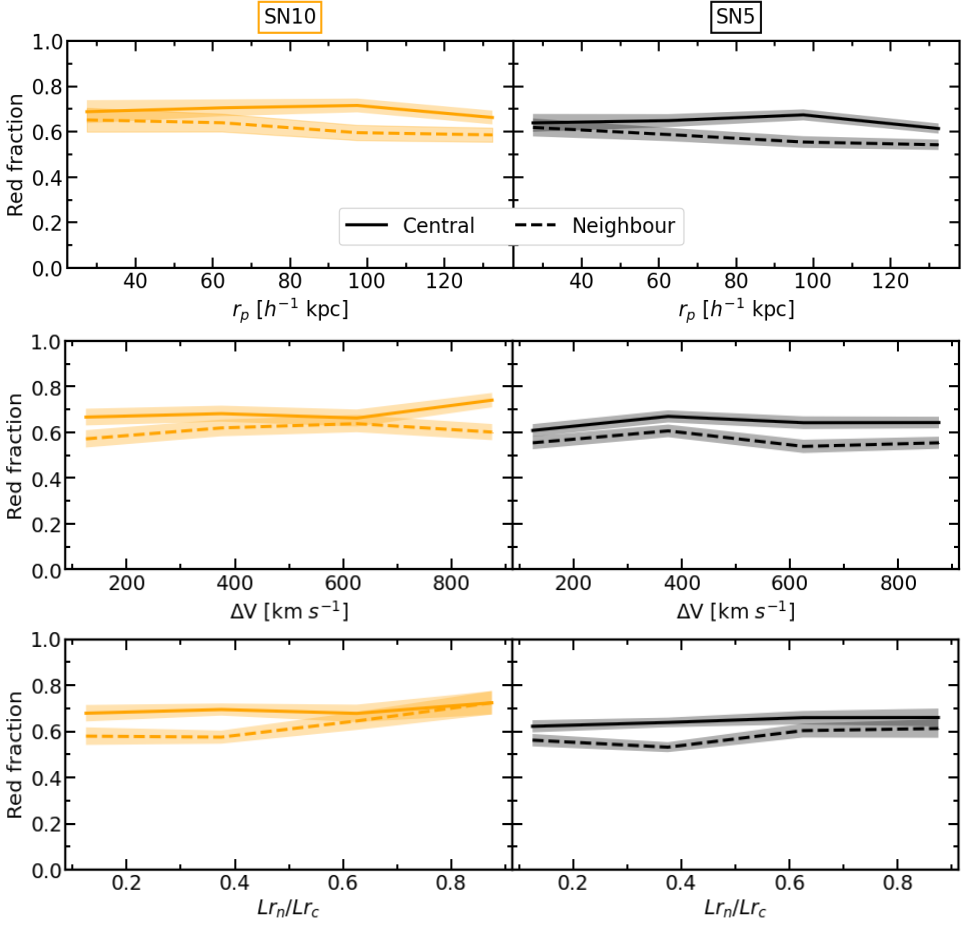}
    \caption{Fraction of red galaxies in the pair sample, for both neighbour (dashed line) and central galaxies (solid line) for both SN10 (gold, left panels) and SN5 (black, right panels) samples, as a function of different properties. \textit{Top panel}: as a function of projected distance. \textit{Middle panel}: as a function of velocity difference. \textit{Bottom panel}: as a function of $r$-band luminosity ratio. Errors were calculated using the bootstrap technique.} 
    \label{fig:froj}
\end{figure*}

Figure \ref{fig:froj} shows the fraction of red neighbours (dashed line) and red centrals (solid line) in gold and black for SN10 and SN5 respectively. They are shown in each panel as a function of projected distance (top), velocity difference (middle) and luminosity ratio (bottom). The stated errors have been computed through the bootstrap technique using $n=20000$ re-samples. 
In all cases we can see that the fractions are above 50$\%$, indicating that most of the pairs are formed by red galaxies independently of their $r_p$, $\Delta V$ and $Lr_n/Lr_c$.

Top and middle panels of Fig. \ref{fig:froj} show the red fractions vs. $r_p$ and $\Delta V$, respectively.
As expected, the red fractions are higher for central galaxies, because they are selected at brighter magnitudes, implying in general redder colours.
Both SN10 and SN5 samples show a relatively stable red fraction across the entire range of projected distances and velocity differences, with fractions generally ranging between $55\%$ and $75\%$. This aligns with the findings of \cite{Patton2011}, who also observed a limited variation in the red fraction of paired galaxies with increasing projected distances, with values around $60\%$.
However, an interesting trend emerges when we consider the difference between the red fractions of central and neighbour galaxies. As the separation between them increases (represented by larger $r_p$ and $\Delta V$ values), the difference becomes more pronounced. For example, in the SN10 (SN5) sample, the difference is around $4\%$ ($2\%$) for the closest pairs in $r_p$, while it rises to $8\%$ ($7\%$) for the most separated ones. Similar trends are observed with velocity difference.
This suggests that red central and red neighbour fractions become more similar in close pairs compared to wider pairs. This could be due to the influence of the central galaxy on its neighbour. As the neighbour gets closer, it might experience environmental effects from the central galaxy (such as gas stripping, tidal interactions, and gravitational influences) that potentially affect its star formation and ultimately its colour.\\

Bottom panels of Fig. \ref{fig:froj} show the fraction of red galaxies as a function of the ratio of luminosities in the $r$-band between the neighbour and the central. Central red galaxies fraction keeps steady, around $68-72\%$ for SN10 and $62-66\%$ for SN5. 
This suggests that their inherent red colour, likely due to past star formation and higher stellar mass \citep{Baldry2006}, remains stable. Neighbours, on the other hand, exhibit a subtle colour shift with increasing luminosity ratio: at lower ratios, where the central is significantly brighter, the neighboring galaxy tends to be less red than expected. This might suggest that factors beyond the central galaxy's influence, such as the inherent properties of smaller galaxies (e.g., gas richness and blue coloration), play a role in shaping the observed color.
However, as the luminosity ratio approaches parity, the fraction of red neighbours starts to climb, eventually surpassing that of centrals. This trend hints at two potential scenarios: red centrals, with their abundant older stars, might be less susceptible to further reddening via environmental influences.
Bluer neighbours, experiencing ongoing star formation, might gradually become redder as their luminosity increases, catching up with the centrals.

\subsection{Characterisation of the global colour properties of the pairs}
\label{sub:global_colour}

In this subsection, we analyse the global properties of our galaxy pairs. To do this, we divided the pairs of galaxies into three types: red pairs, blue pairs, and mixed pairs, depending on whether both galaxies in the pair are red, both are blue, or one is red and the other is blue, respectively. Table \ref{tab:pares} shows the resulting number of pairs for each type. 

The data presented in Table 2 provide evidence for the phenomenon known as "galactic conformity". This term describes the tendency of galaxies within a group or cluster to share similar properties with the central galaxy. Notably, satellite galaxies often resemble the central galaxy in shape, with central spirals having more spiral satellite galaxies and central ellipticals having more elliptical companions. 
Indeed, our analysis of galaxy pairs reveals a strong association between the colour of a galaxy and its companion. 
Within galaxy pairs, the chances of finding both galaxies sharing the same colour are higher than finding a mixed. This pattern holds true for both SN10 and SN5 samples. Specifically, when one galaxy in a pair is red, its companion has a $46\%$ (for SN10) or $39\%$ (for SN5) chance of also being red. Similarly, if one galaxy is blue, the probability its partner is also blue is $16\%$ (for SN10) or $19\%$ (for SN5). Conversely, only $38\%$ for SN10 and $42\%$ for SN5 are mixed pairs.
This significant correlation observed between the colours of galaxy pairs provides additional support for the existence of galactic conformity and suggests a shared evolutionary history for galaxies within close proximity.\\

\begin{table}
\begin{center}
\begin{tabular}{|c|c|c|c|c|c|c|}
\hline
\multirow{2}{*}{} & \multirow{2}{*}{Total pairs} & \multicolumn{2}{ c|}{Galaxies in pairs} & \multicolumn{3}{ c|}{Pairs types} \\ \cline{3-7}
 &  & Red & Blue & Red & Blue & Mixed\\ \hline
SN10 & 700 & 908 & 492 & 321 & 113 & 266 \\
SN5 & 1278 & 1539 & 1017 & 500 & 239 & 539 \\ \hline
\end{tabular}
\caption{SN10 and SN5 isolated galaxy pairs and their colour classification. The first column shows the total number of pairs in each sample; the second column shows the number of red and blue galaxies in the pairs; and the third column the number of each pair type: red pairs (formed by two red galaxies), blue pairs (by two blue galaxies), and mixed pairs (by one red and one blue galaxy). }
\label{tab:pares}
\end{center}
\end{table}

\begin{figure*}
	\includegraphics[width=1.5\columnwidth]{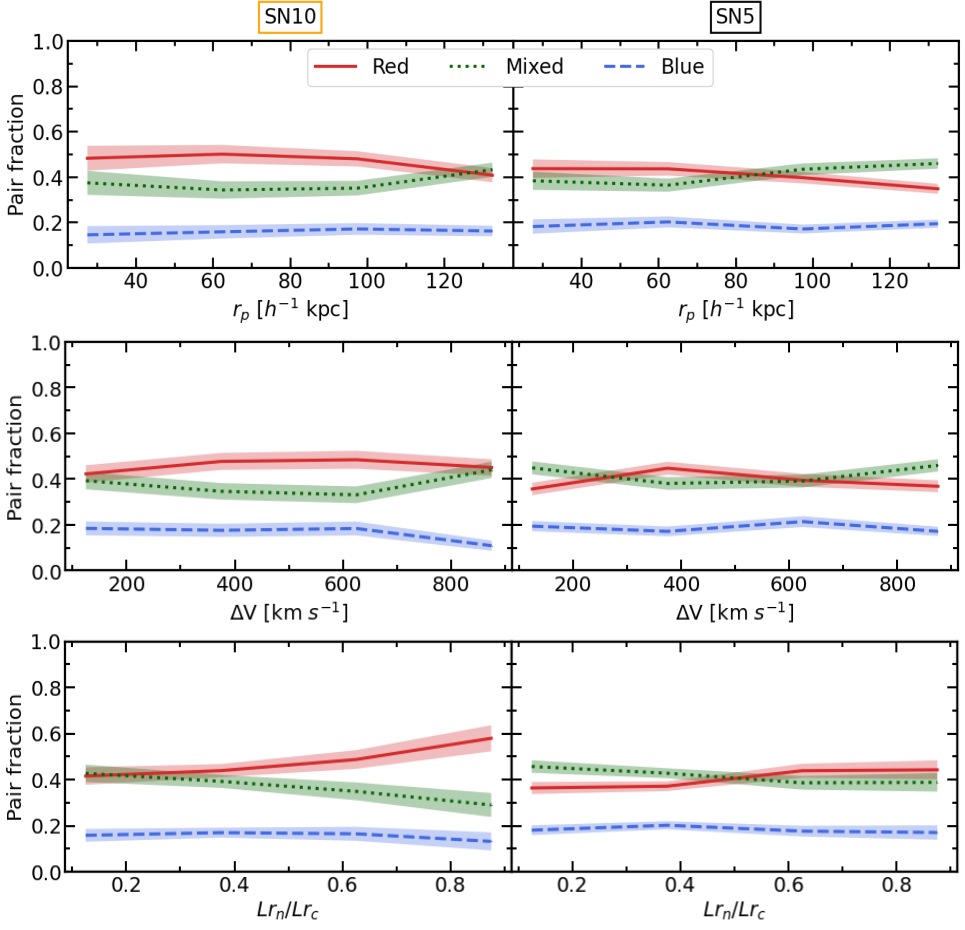}
    \caption{Fraction of each type of pair: red pairs (solid red lines), mixed pairs (green dotted lines), and blue pairs (blue dashed lines) for SN10 (\textit{left panels}) and SN5 (\textit{right panels}) samples, as a function of different properties. \textit{Top panels}: as a function of projected distance. \textit{Middle panels}: as a function of velocity difference. \textit{Bottom panels}: as a function of $r$-band luminosity ratio between the neighbour and central galaxy. Errors were calculated using the bootstrap technique.} 
    \label{fig:fpar}
\end{figure*}

Delving deeper into the composition of galaxy pairs, we investigated how the fractions of red, mixed, and blue pairs vary with different properties. Figure \ref{fig:fpar} visually depicts these relationships for both SN10 and SN5 samples. Each colour (red, green, and blue) corresponds to the fraction of its respective pair type. Notably, we used the bootstrap technique to ensure accurate error bars for each category.

In Figure \ref{fig:fpar}, the blue pairs maintain constant fractions at approximately $\sim 14-17 \%$ (SN10) and $\sim 17-20 \%$ (SN5), as their general behaviour. This indicates that the blue fractions are independent of $r_p$, $\Delta V$, and $Lr_n/Lr_c$.
One notable point is that for larger velocity differences, the fraction of blue pairs drops to 10$\%$ (SN10), possibly due to the absence of pairs at these values that trigger bursts of star formation.
On the other hand, both red and mixed pairs have higher fractions than blue pairs, and vary in both samples and in all properties. 
For the case of SN10, the fraction of red pairs is constant at smaller $r_p$ ($\sim 50\%$), but decreases at higher $r_p$ ($\sim 40\%$) as the mixed pairs take a higher fraction (from $ 34\%$ to $ 43\%$).
This indicates that as the pair members are closer together, there is a higher probability that they are both red, whereas as they move further apart, it becomes more likely that they are mixed.
This is consistent with the previous section.
As a function of the velocity difference, the fractions tend to remain constant between $ 42\%$ and $ 48\%$, being red pairs more frequent than mixed pairs.

As a function of the luminosity ratio, Fig. \ref{fig:fpar} shows that the fraction of red pairs increases with the ratio from $\sim$0.4 to 0.6, and mixed pairs decrease from $\sim$0.4 to 0.3 for SN10. We can note that when 
the galaxies have similar luminosities, most of the pairs tend to be formed by two red galaxies ($\sim 60 \%$). However, when the luminosities of the members differ greatly, the fraction of red pairs decreases slightly, and mixed pairs are frequent too.
This suggests that similar luminosities might favor the formation of pairs comprised of two red galaxies. Conversely, when the luminosities 
diverge significantly, the likelihood of finding mixed pairs increases. 
The SN5 sample displays similar overall trends, as shown in the right panel of Figure \ref{fig:fpar}. However, slight variations compared to SN10 could be attributed to the lower signal-to-noise ratio inherent to its image data.

\subsection{Galaxy pair environments} \label{densidad}

\begin{figure*}
    \centering
    \includegraphics[width=1.9\columnwidth]{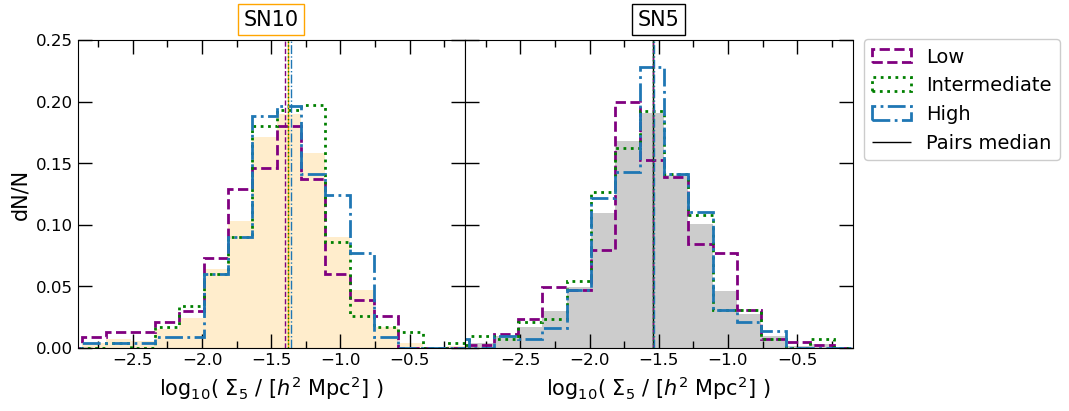}
    \caption{Distributions of the projected local density $(\Sigma_5)$ for the SN10 (gold shaded, \textit{left panel}) and SN5 (black shaded, \textit{right panel}) galaxy pairs. The distributions for the low, intermediate, and high stellar mass samples are shown in dashed, dotted, and dash-dotted lines, respectively. Vertical lines show the medians of the samples.}
    \label{fig:sigma5_masas}
\end{figure*}

To characterise the environments of galaxy pairs, we computed the projected galaxy densities using the nearest neighbours in the plane of the sky. 
Following \cite{Balogh2004} and \cite{Baldry2004b}, we used the area defined by the 5th nearest neighbour to each central galaxy (excluding its pair companion), brighter than $M_{r} -5log(h) \leq$ -20.5 and within $\Delta V <$ 1000 km s$^{-1}$ to estimate the projected local density ($\Sigma_5$).
The local density estimator is defined as $\Sigma_5=5 / (\pi r_5^2) $, where $r_5$ is the projected distance from the central galaxy to the 5th nearest neighbour galaxy, and is measured in $h^{-1}$Mpc$^{-2}$.
The shaded histograms in Figure \ref{fig:sigma5_masas} show the distributions of the projected local density for both pairs samples. They are mostly concentrated between -2 and -1, reaching a median value of log$_{10}(\Sigma_5) = -1.38$ for SN10 and log$_{10}(\Sigma_5) = -1.54$ for SN5 (vertical lines).  
These findings show lower densities than other studies, such as those conducted by \cite{Perez2009,Lambas2012,Mesa2014}. 
This may be a result of our isolation criteria, in which we are asking for our pairs to be separated by more than 450 $~\mathrm{kpc}$ and 1000 $~\mathrm{km\,s}^{-1}$ from other structures. In particular, these mentioned works do not consider any isolation criteria.
On the other hand, it is well-known that the fraction of red galaxies increases with density \citep{Balogh2004,Baldry2006,Cooper2006}. The fact that our pairs are in low and intermediate-density environments and have a high fraction of red galaxies, 
this is further implied by the pre-processing these systems may have undergone, potentially shaping their current state and influencing the observed trends.

In order to study galaxy pairs that reside in the same local-density regions and have similar stellar mass, we analysed three different ranges of stellar mass: low, intermediate, and high. 
Using the stellar masses calculated in Section \ref{mass}, we divided the pairs according to the mass of the pair using the 33.33rd and 66.66th percentiles of the distributions. For the SN10 sample, the ranges are defined by log$_{10}$(M$_*/{\rm M}_\odot)=$ 10.85 and 11.21. In the case of SN5 sample, the percentiles are log$_{10}$(M$_*/{\rm M}_\odot)=$ 10.78 and 11.16. 
By choosing our analysis by stellar mass, we effectively minimize the influence of confounding factors like environment and differences in galaxy size. This allows us to isolate and examine potential interactions between galaxies within each mass category, unmasked by environmental bias or mass discrepancies. 
Fig. \ref{fig:sigma5_masas} confirms this independence visually and, by performing a Kolmogorov-Smirnov (KS) test, we found that the chosen stellar mass cuts (represented by different dashed lines) show no relation to the projected local density distribution. \\

\begin{figure*}
    \centering
    \includegraphics[width=1.7\columnwidth]{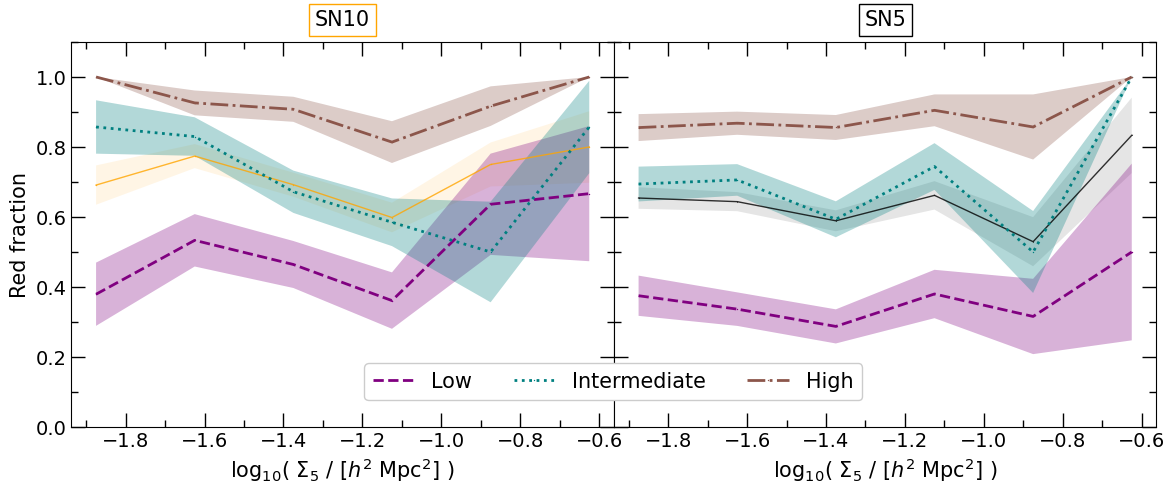}
    \caption{Fraction of red central galaxies as a function of projected local density ($\Sigma_5$) in the SN10 sample (solid gold line, \textit{left panel}) and SN5 sample (solid black line, \textit{right panel}). The red fractions for the low, intermediate, and high stellar mass samples are shown in dashed, dotted, and dash-dotted lines, respectively. Errors are calculated using the bootstrap technique. }
    \label{fig:froj_masa}
\end{figure*}

Figure \ref{fig:froj_masa} shows the fraction of central red galaxies as a function of the projected density for both SN10 pairs in the left panel and SN5 pairs in the right panel. Solid lines show full samples, whereas the low stellar mass samples are shown in dashed lines, the intermediate stellar mass in dotted lines, and the high stellar mass in dash-dotted lines.
In both pairs samples, it is noticeable that the fraction of red centrals exceeds $50\%$ across the entire range, showing a slight increase towards higher densities as expected.
\cite{Perez2009} found that at intermediate densities (in the range of our higher densities), the red fraction of galaxies in close pairs exceeds that of isolated galaxies, suggesting that in intermediate-density environments galaxies are efficiently pre-processed \citep[e.g.][]{zabludoff98,fujita2004} by close encounters and mergers before entering higher local density regions.
Even though our sample encompasses close and wide pairs, the red fractions also increase at higher densities but show a difference according to their stellar mass.

For high stellar masses, the red fraction is the highest and appears independent of density. This suggests that environmental factors might play a less significant role in shaping the colour of these massive galaxies.
At lower stellar masses, a different pattern emerges. The red fraction remains relatively constant across lower densities but increases as the density rises. This could indicate that denser environments preferentially host red central galaxies in this mass range.
Intermediate stellar masses present a more nuanced picture. Here, the red fraction seems to dip in intermediate-density regions before gradually climbing again at higher densities. However, the noise is higher in this interval of projected density.
Overall, these findings highlight the complex interplay between galaxy colour, stellar mass, and local environment. \\

\begin{figure*}
    \centering
    \includegraphics[width=1.7\columnwidth]{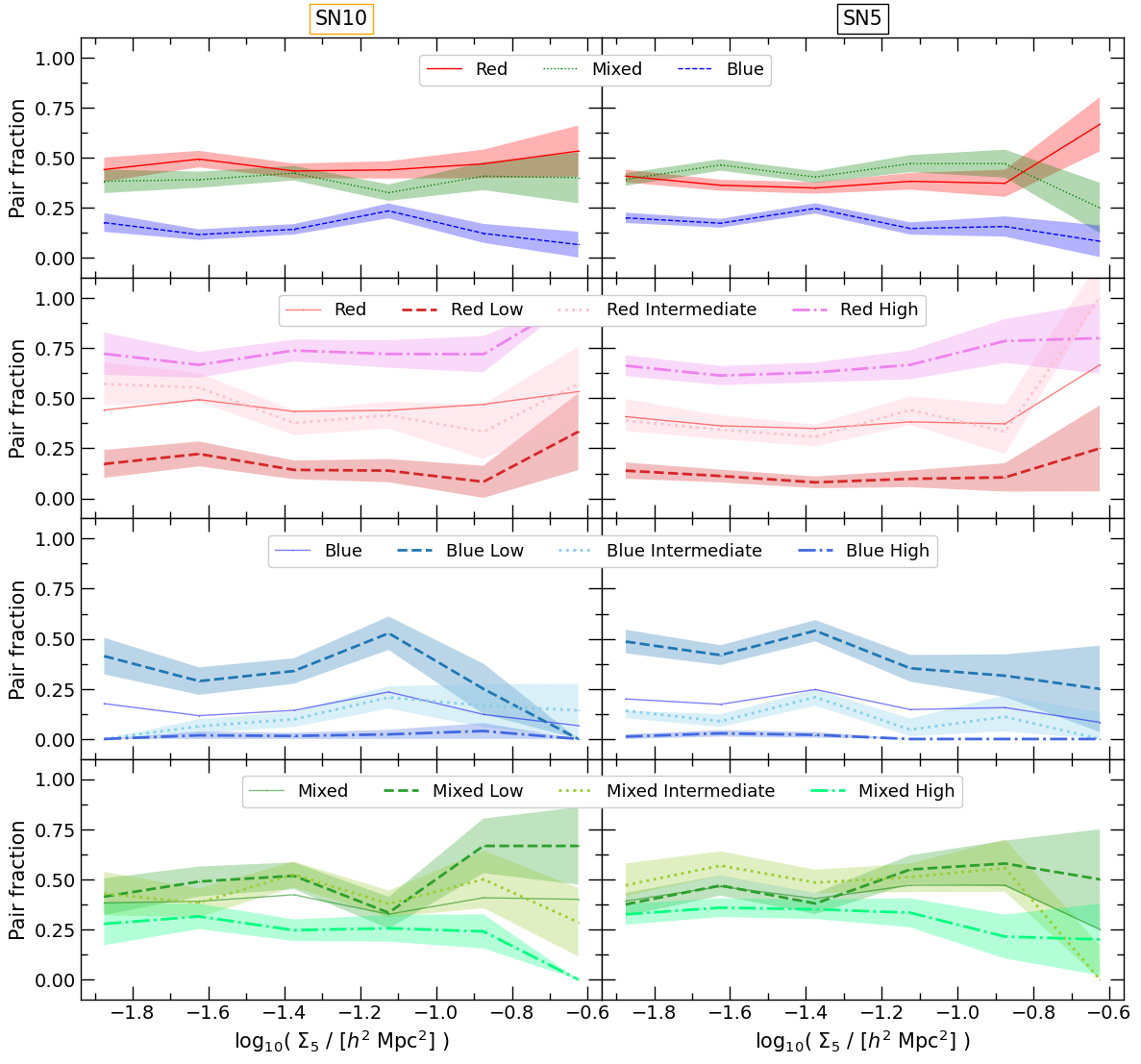}
    \caption{ Fraction of red pairs (red and pink colours), blue pairs (blue and light blue colours) and mixed pairs (green colours), for both SN10 sample (\textit{left panels}) and SN5 (\textit{right panels}), as a function of projected local density.
    In the first panel, the full sample is shown and in the rest, it is divided into high, intermediate, and low stellar mass samples.
    Errors are calculated using the bootstrap technique. }
    \label{fig:frac_masa}
\end{figure*}

The Figure \ref{fig:frac_masa} shows the fraction of the different types of pairs (red, blue, and mixed pairs) defined in Sec. \ref{sub:global_colour} as a function of the projected local density. 
The top panels show the three fractions of galaxy pairs for the entire stellar mass range as a function of $\Sigma_5$. 
The red and mixed galaxy pairs seem to be independent of the environment, showing a similar fraction across the entire density range. 
In the SN5 sample, we noted an increase in the fraction of red pairs in higher-density bins (from 37$\%$ to 67$\%$) and a decrease in the fraction of mixed (from 47$\%$ to 25$\%$). 
Although this is not evident in the entire SN10 sample, the trends become noticeable when are analysed in stellar mass slices, suggesting that the red pair fraction is higher in high-density environments and the mixed fraction depends on its stellar mass. 

In the case of the red pairs, it is evident that they are primarily composed of those with the higher stellar mass, followed by those of intermediate mass, and in the minority, by those of lower mass. This reflects the relationship between stellar mass and the colours of galaxies.
In all three cases, the fraction of red pairs increases at higher projected densities, indicating that the environment acts independently of mass.

On the other hand, for mixed pairs, the differences between the fractions of different stellar masses are less pronounced, but they all maintain a similar behaviour at lower densities.
In denser environments, an increase in the fraction of low-mass mixed pairs is noticeable, whereas, for the more massive mixed pairs, the fractions decrease.
This indicates that there are no high stellar-mass mixed pairs in dense environments, which may be due to the fact that the red galaxies dominate those systems. This is possibly because there are no high-mass pairs composed of one red and one blue galaxy, as typically observed are the presence of a massive red and a less massive blue galaxy.

Finally, in the case of blue pairs, the general fractions fall to $\sim 8\%$ in environments with higher densities, indicating that they are not immersed in larger systems.
In particular, the fraction of high-mass blue pairs is practically zero and independent of density. This shows that blue pairs are not formed by massive galaxies. However, for lower masses, the blue pairs fraction is higher at low densities and decreases with density, could be interpreted as reflecting environmental influences and pre-processing.
It can be clearly seen that most of the blue pairs are formed by low-mass central galaxies, while the red pairs are formed by higher-mass central galaxies.

\section{Conclusions} \label{sec:conclusions}

In this work, we present the first catalogue of isolated galaxy pairs for the S-PLUS DR4 and a characterization of their properties. Here, we present a summary and conclusions of this study. 

To obtain the sample of isolated galaxy pairs, we applied an algorithm to identify them in two complete S-PLUS flux samples, one with a signal-to-noise ratio greater than 10 (SN10 sample) and the other greater than 5 (SN5 sample). The algorithm follows a traditional approach and it is based on two parameters: the projected distance ($r_p$) and the velocity difference ($\Delta V$), which were tested on a simulated catalogue that reproduces the characteristics of the S-PLUS. 
From the analysis of the completeness and purity of the pairs, we chose the values of $r_{p,max}$ = 150 $h^{-1}$ kpc and $\Delta V_{max}$ = 1000 km s$^{-1}$ and an isolation criterion of $3r_{p,max}$ to identify them. 
We chose those parameters such that they imply a completeness of $\sim 77\%$ and a contamination of less than $\sim 20\%$. 
With them, we obtained 1278 and 700 isolated galaxy pairs for SN5 and SN10, respectively.

The precision of the S-PLUS DR4 photometry provides us the opportunity to delve deeper into the characteristics of our galaxy pairs. This enables us to explore various properties, such as their colours and stellar masses, and to determine their luminosity function, all calculated through photometric techniques.
Our results fall within the range established by other studies, further validating the accuracy and reliability of our analysis.
In particular, the parameters of our luminosity function of galaxies in pairs (characteristic luminosity and faint-end slope) align consistently with those of other authors who used a similar photometric band and calculated the LF in small groups and loose/poor galaxy clusters. This suggests that the luminosity function might be a characteristic feature of galaxy systems.

Driven by the notion that galaxy pairs are systems hosting higher fractions of red galaxies compared to isolated galaxies, one of our main objectives is to understand how this correlates with the pair's characteristics (projected distance, velocity difference, luminosity ratio, and stellar mass) and their local environment.

One of the results we obtained is that independently of $r_p$, $\Delta V$ and luminosity ratio, our samples show a high fraction of red galaxies in all ranges.
Analysing the fraction of red galaxies, for both the central members and their neighbours (Fig. \ref{fig:froj}), we found evidence that as the neighbour gets closer to its central in $r_p$ and $\Delta V$, it might be subjected to environmental effects from the central. The gas stripping, tidal interactions, and gravitational influences – all potentially affecting the neighbour's star formation and ultimately, its colour.

Analysing the relationship between luminosity and colour, we observe contrasting stories for central and neighbour galaxies. 
Central galaxies, maintain a stable red colour regardless of luminosity. This likely reflects their reddish, possibly due to past starbursts and their stellar mass. 
Neighbours, on the other hand, tell a more nuanced tale. 
When the central galaxy is significantly brighter, the neighbour tends to be less red. This could hint at the influence of the central galaxy's surroundings on its neighbour's star formation. However, as the luminosity ratio approaches parity, the red fraction of neighbours starts to climb, eventually surpassing that of the central ones.
This trend presents two plausible scenarios:
Red centrals, with older stars, might be less susceptible to further reddening from environmental influences. The bluer neighbours, still actively forming stars, could gradually become redder as their luminosity increases, eventually catching up with their central counterparts.

Driven by the initial findings, we categorized the pairs into three distinct groups: red pairs, blue pairs, and mixed pairs. This allows us to delve deeper into the colour conformity observed within these systems. Within each pair, the chances of finding both galaxies sharing the same colour are significantly higher than encountering a mixed pair. This  suggests that galaxies residing in close proximity might share a common evolutionary history shaped by similar environmental factors or even direct interactions.

To understand and deepen this connection, we investigate the fractions of these pairs (red, mixed and blue) across different properties. 
As illustrated in Fig. \ref{fig:fpar}, 
blue pairs fractions are in general independent of $r_p$, $\Delta V$ and $Lr_n/Lr_c$, but as the pair members are closer together, there is a higher probability that they are both red. 
Our finding also reveals a relationship between luminosity and colour pairing: when the luminosities of the galaxies in a pair are similar, the likelihood of encountering a red pair significantly increases. This suggests that shared environments or interactions might be fostering the formation of red galaxies in close proximity. Conversely, when the luminosities of the pair members diverge, the probability of encountering a mixed pair (one red, one blue) rises. This could indicate that environmental influences or interactions have a less pronounced effect on galaxies with disparate luminosities, allowing for the coexistence of both red and blue galaxies within a pair. \\

Analysing local density, stellar mass, and photometric properties, we discovered that environmental factors play a less significant role in shaping the colour of massive central galaxies. This suggests that their red hue might be more deeply ingrained, perhaps a result of past star formation or internal dynamics. However, for central galaxies with lower stellar masses, that denser environments preferentially host red galaxies in this mass range. 

The expected colour-density relation is of course recovered, but it is strongly determined by the stellar mass of the pair.
Whereas as the density increases the fraction of red galaxies increases and the fraction of blue pairs decreases, the fraction of mixed pairs is determined by their stellar mass. In denser environments, the fraction of more massive mixed pairs decreases, while that of less massive pairs increases. This is related to the fact that there are no high stellar-mass mixed pairs in dense environments and in general the blue pairs are formed by low-mass central galaxies, while the red pairs are formed by higher-stellar mass galaxies. 

It is interesting to put these results in the context of the so-called pre-processing, which suggests that galaxy evolution is affected by the environment even before galaxies arrive in clusters. Indeed, the large fraction of red pairs in our sample provides evidence that even in small systems environmental factors, such as the suppression of star formation, play an important role.

This work presents a significant contribution to the study of galaxy pairs by applying our algorithm to the S-PLUS survey. This process has resulted in the creation of reliable galaxy pair catalogues, which are now publicly available online \footnote[4]{  \url{https://catalogs.iate.conicet.unc.edu.ar/pairs_splus+dr4/}}. 
These catalogues offer a wealth of information for researchers, including the main photometric properties of each galaxy in the pair. See Appendix \ref{apend:sample} for a complete list of published catalogue information.
By providing these resources, we aim to facilitate further research and exploration in the field of galaxy interactions and their impact on galactic evolution.

\section*{Acknowledgements}

This work has been partially supported with grants from Agencia Nacional de Promoci\'on Cient\'ifica y Tecnológica, the Consejo Nacional de Investigaciones Cient\'{\i}ficas y T\'ecnicas (CONICET, Argentina) and the Secretar\'{\i}a de Ciencia y Tecnolog\'{\i}a de la Universidad Nacional de C\'ordoba (SeCyT-UNC, Argentina).
This work has been partially supported with grants from funding of the grant 2019/26492-3, which is the gant that funds T80-South at the moment.
MCC acknowledges the support from Beca EVC-CIN 2020.
FR would like to acknowledge support from the ICTP through the Junior Associates Programme 2023-2028.
LSJ acknowledges the support from CNPq (308994/2021-3)  and FAPESP (2011/51680-6).
The authors appreciate the helpful comments and suggestions made by the anonymous referee, which improved this article.

\section*{Data Availability}
The data underlying this article will be shared on reasonable request to the corresponding authors. 
The  pairs catalogues are available online.
See Appendix \ref{apend:sample} for a complete list of published catalogue information.



\bibliographystyle{mnras}
\bibliography{example} 



\appendix

\section{Published pair sample} \label{apend:sample}

The S-PLUS DR4 galaxy pairs catalogues were compiled for a sample of galaxies with signal-to-noise 10 and another with signal-to-noise 5.
They are available at \url{https://catalogs.iate.conicet.unc.edu.ar/pairs_splus+dr4/} and contain the following columns.

$ID\_par$: Galaxy pair identification 

$ID\_gal$: Galaxy identification

$ra [^{\circ}]$: Right Ascension

$dec [^{\circ}]$: Declination

$zp$: Photometric redshift

$petro\_mag\_r [mag]$: r-Petrosian band magnitude

$petro\_mag\_g [mag]$: g-Petrosian band magnitude

$rp [kpc h^{-1}]$: Projected distance between the central galaxy and its neighbour. If rp=0 it indicates that it is the central galaxy

$star\_mass [M\odot h^{-2}]$: Stellar mass estimated following Yang et al. 2007

$SN$: Signal-to-noise of the sample it was constructed from

\bsp	
\label{lastpage}
\end{document}